\documentclass[12pt,draftclsnofoot,onecolumn]{IEEEtran}
\usepackage{amsmath}
\usepackage[dvips]{graphicx}
\DeclareGraphicsExtensions{.eps}

\usepackage{amssymb}
\usepackage{array}
\usepackage{cite}
\begin{document}

\title{Cluster-based Transform Domain Communication Systems for High Spectrum Efficiency}

\author{\normalsize Su~HU$^{\dag}$$^{\ddag}$,
        Yong Liang~GUAN$^{\dag}$,
        Guoan~BI$^{\dag}$,
and
        Shaoqian LI$^{\ddag}$\\
        {{$^{\ddag}$ National Key Laboratory of Science and Technology on Communications, University of
Electronic Science and Technology of China, 2006 Xiyuan Avenue, Chengdu, China, 611731} \\
$^{\dag}$ School of EEE, Nanyang Technological University, 50 Nanyang Avenue, Singapore, 637553}\\{E-mail: husu@uestc.edu.cn} }

\maketitle

\begin{abstract}
This paper presents a cluster-based transform domain communication system (TDCS) to improve
spectrum efficiency. Unlike the utilities of clusters in orthogonal frequency division multiplex
(OFDM) systems, the cluster-based TDCS framework divides entire unoccupied spectrum bins into $L$
clusters, where each one represents a data steam independently, to achieve $L$ times of spectrum
efficiency compared to that of the traditional one. Among various schemes of spectrum bin spacing
and allocation, the TDCS with random allocation scheme appears to be an ideal candidate to
significantly improve spectrum efficiency without seriously degrading power efficiency. In
multipath fading channel, the coded TDCS with random allocation scheme achieves robust BER
performance due to a large degree of frequency diversity. Furthermore, our study shows that the
smaller spectrum bin spacing should be configured for the cluster-based TDCS to achieve higher
spectrum efficiency and more robust BER performance.
\end{abstract}


\section{Introduction}

Due to the scarcity of available spectrum, future wireless communication systems have to
efficiently use all available spectrum resources. The concept of transform domain communication
system (TDCS) has been initially studied in \cite{Andren}, where it smartly synthesizes an
adaptive waveform by avoiding using spectrum bands occupied by jammers or licensed users. Unlike
orthogonal frequency division multiplexing (OFDM) and multi-carrier code division multiple access
(MC-CDMA), TDCS is designed to avoid the use of occupied bands by signal processing facilities at
both transmitter and receiver instead of mitigating the interference only at receiver
\cite{Chakravarthy-TDCS}. Thus, TDCS can be used as a cognitive radio (CR) modulation technique
for overlay opportunistic spectrum access systems \cite{Budiarjo}\cite{Chakravarthy2009}.

The multiple access TDCS (MA-TDCS) has been first implemented by assigning each user a unique
primitive polynomial for a different $m$-sequence \cite{Swackhammer}, and the techniques needed
for acquisition and synchronization have also been discussed in \cite{Roberts2000}. The authors in
\cite{HanC} have proposed an efficient implementation of TDCS to enhance bit error rate (BER)
performance by removing the imaginary part of noise components. For practical applications, the
problem of peak-to-average power ratio (PAPR) has been studied to minimize the nonlinear
distortion of high power amplifier (HPA) \cite{Martin}. However, TDCS has been used only for
the low-rate control channel in cognitive radio networks because of its low spectrum efficiency
\cite{Han}. In order to improve spectrum efficiency, a modified TDCS model has been proposed with
another data source in the form of embedded symbols \cite{Budiarjoembed}. Since it uses the
high-order phase shift keying (PSK) modulation, the embedded TDCS achieves unsatisfactory spectrum
and power efficiency due to the degraded BER performance with reduced Euclidean distance.

In OFDM systems, the concept of clustering has been widely used for channel estimation
\cite{LiCluster} or interference suppression \cite{ZhangCluster}. However, the purpose of using
clusters is not to improve spectrum efficiency since OFDM is essentially a digital modulation
technique where the data stream modulates spectrum bins directly. In this paper, we present a
cluster-based TDCS framework to improve spectrum efficiency, based on the fact that it still
achieves sufficient orthogonality among different spectrum bins when the number of unoccupied
spectrum bins is smaller than the order of cyclic code shift keying (CCSK) modulation
\cite{Fumat}. For the $L$-cluster TDCS, the unoccupied spectrum bins are divided into $L$
clusters. We will show that our proposed TDCS achieves $L$ times of spectrum efficiency compared
to that of the traditional one.

Similar to many communication scenarios, the cluster-based TDCS encounters the tradeoff between
spectrum and power efficiency, i.e., it achieves high spectrum efficiency with a penalty of BER
performance as the number of clusters increases \cite{Han}. To solve this problem, in this paper,
two spectrum bin allocation schemes are considered, namely the continuous and random allocation
schemes. Analytical and simulation results illustrate that, the $L$-cluster TDCS with random
allocation scheme achieves $L$ times of spectrum efficiency without serious BER performance
degradation, compared to the traditional one. We also find that, different from the result in
\cite{Fumat}, the spectrum bin spacing of the cluster-based TDCS should be configured as small as
possible to achieve high spectrum efficiency and robust BER performance.

We use $\left( \cdot \right)^* $ and $| \cdot |$ to represent the operations of conjugate and
absolute value. The modulo operation $mod \left( x, y \right)$ denotes the remainder of $x$
divided by $y$. The symbols $\mathcal {F}$ and $\mathcal {F}^{-1}$ represent the operations of
fast Fourier transform (FFT) and its inverse (IFFT), respectively. Finally, the symbol $\emptyset$
denotes an empty set.

\section{Review on Traditional TDCSs}
\begin{figure}
\centering
  \includegraphics[width=4.2in]{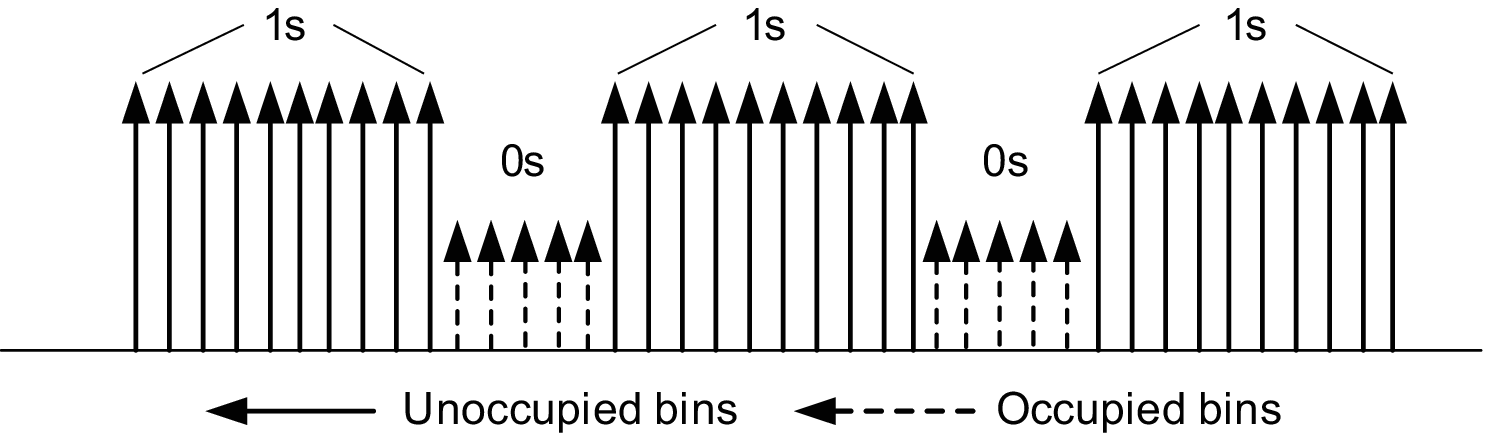}
  \caption{White/black space spectrum and spectrum availability vector $\mathbf{{A}}$}
  \label{FIG:SpecIdentify}
\end{figure}

\begin{figure}
\centering
  \includegraphics[width=4.2in]{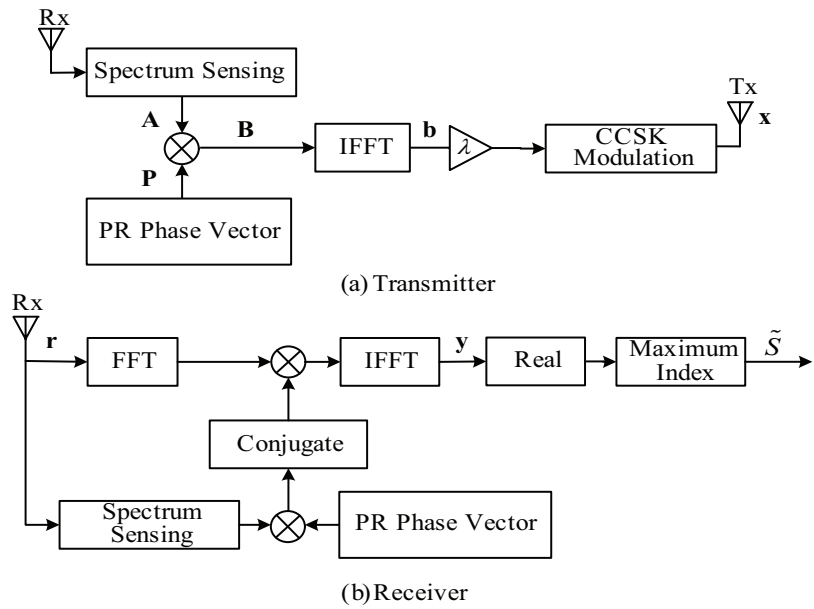}
  \caption{Block diagram of the traditional TDCS}
  \label{FIG:TraTDCSDiagram}
\end{figure}

In TDCS, the entire spectrum band is divided into $N$
spectrum bins. A spectrum availability vector, $ {\mathbf{{A}}} = \left\{ {A_0 ,A_1 ,...,A_{N
- 1} } \right\}$, is used to represent the distribution of spectrum holes as shown in Fig.\ref{FIG:SpecIdentify}. Note that the value of $A_k$ is set to 1 (or 0) if the $k_{th}$ bin is unoccupied (or occupied). Let us assume that there are $N_C$ unoccupied bins inside the set $\Omega ^C$, i.e.,
$\left\{ {A_k = 1,k \in \Omega ^C } \right\}$ \cite{Chakravarthy-TDCS}. According to Fig.\ref{FIG:TraTDCSDiagram}, a user-specific complex pseudorandom (PR) phase vector, ${\mathbf{{P}}} = \{ e^{jm_0 }, e^{jm_1 }, ..., e^{jm_{N - 1} } \}$, is multiplied element by
element with $\mathbf{{A}}$ to produce a spectral vector $\mathbf{{B}}$, i.e., $\mathbf{B}=\mathbf{A} \cdot \mathbf{P}$. The fundamental modulation
waveform (FMW) $\mathbf{{b}}$ is achieved by performing an IFFT operation,
\begin{equation}
\label{b-FMW} {\mathbf{{b}}}= \left\{ {b_0 ,b_1 ,...,b_{N - 1}} \right\} = \lambda \mathcal
{F}^{-1} \{ \mathbf{{B}} \},
\end{equation}
where $\lambda  = \sqrt {{N / N_C }}$ is an energy normalization factor. With a $M$-ary CCSK modulator, the
transmitted waveform, ${\mathbf{{x}}}= \left\{x_0, x_1, \cdots, x_{N-1} \right\}$, is achieved by cyclically shifting $\mathbf{b}$ with $S$ places \cite{Han},
\begin{equation}\label{Trad}
x_n  = b_{mod \left( n-\frac{SN}{M}, N \right)} = \lambda {\sum\limits_{k = 0}^{N - 1} {A_k
e^{jm_k } e^{\frac{{ - j2\pi S k}}{M}} e^{\frac{{j2\pi kn}}{N}} } }.
\end{equation}

For detection, the received waveform, ${\mathbf{{r}}} =
\left\{r_0, r_1, \cdots, r_{N-1}\right\}$, is correlated with the local reference FMW to recover
input data symbols by detecting the maximum correlation output \cite{Dillard}. To halve noise
effects, the receiver extracts only the real part of maximum correlation output,
\begin{equation}\label{EqCD}
\tilde{S} = \arg \max \left\{ \Re \left\{ \mathcal {F}^{-1} \left\{ \mathcal {F}
(\mathbf{{r}}) \cdot \left( \mathcal {F} (\mathbf{{b}}) \right)^* \right\} \right\}
\right\},
\end{equation}
where $\Re \left\{ \cdot  \right\}$  denotes the operator obtaining the real part of a complex
quantity.

Since each transmitted waveform carries $\log_2 M$ bits, the spectrum efficiency of traditional
TDCS with bandwidth $W$ and spectrum bin spacing $\Delta_f$, i.e., $\Delta_f={W}/{N}$, is
given by \cite{Fumat}
\begin{equation}\label{EffTr}
\eta _{TDCS} = \frac {\Delta _f \log _2 (M)}{\gamma W} \text{(bits/s/Hz)},
\end{equation}
where $\gamma$ denotes the unoccupied bandwidth ratio. According to \eqref{EffTr}, the
traditional TDCS should choose $M$ and $\Delta_f$ with the highest possible value to improve
spectrum efficiency. However, it is emphasized in \cite{Fumat} that $\Delta_f$
should be configured as small as possible to achieve noise-like properties for robust BER
performance. Obviously, this inherent tradeoff with respect to $\Delta_f$ makes it difficult to
achieve robust BER performance and high spectrum efficiency simultaneously.

\section{Cluster-based TDCS}
From \cite{Fumat}, TDCS still achieves sufficient orthogonality among different spectrum
bins when the number of unoccupied spectrum bins is smaller than the CCSK modulation order.
Therefore, in this paper, a cluster-based TDCS framework is proposed to achieve high spectrum
efficiency by dividing the unoccupied spectrum bins into clusters.

We assume that $N_C$ unoccupied spectrum bins are equally divided into $L$ clusters, and each
cluster has $N_C/L$ unoccupied spectrum bins. For the $l_{th}$ cluster, the unique spectrum
availability vector is defined as ${\mathbf{{A}}}^l = \left\{ {A_0^l ,A_1^l , \ldots ,A_{N -
1}^l } \right\}$,
\begin{equation}\label{Eq1}
A_k^l  = \left\{ {\begin{array}{*{20}c}
   {1,} & {k \in \Omega _l^C }  \\
   {0,} & {k \in \Omega _l }  \\
\end{array}} \right.,
\end{equation}
where $\Omega _l^C$ and $\Omega _l$ denote the sets of unoccupied and occupied spectrum bins for
the $l_{th}$ cluster, respectively. To fully utilize all available spectrum resources and maintain
orthogonality among different clusters, $\left\{ {\Omega _l^C ,l = 1,2,...,L} \right\}$ should
satisfy
\begin{equation}\label{Eq2}
\begin{array}{*{20}c}
   {\bigcup\limits_{l = 1,2,...,L} {\Omega _l^C }  = \Omega ^C}& \text{and} & {\bigcap\limits_{l = 1,2,...,L} {\Omega _l^C }  = \emptyset }.  \\
\end{array}
\end{equation}

\begin{figure}
\centering
  \includegraphics[width=4.2in]{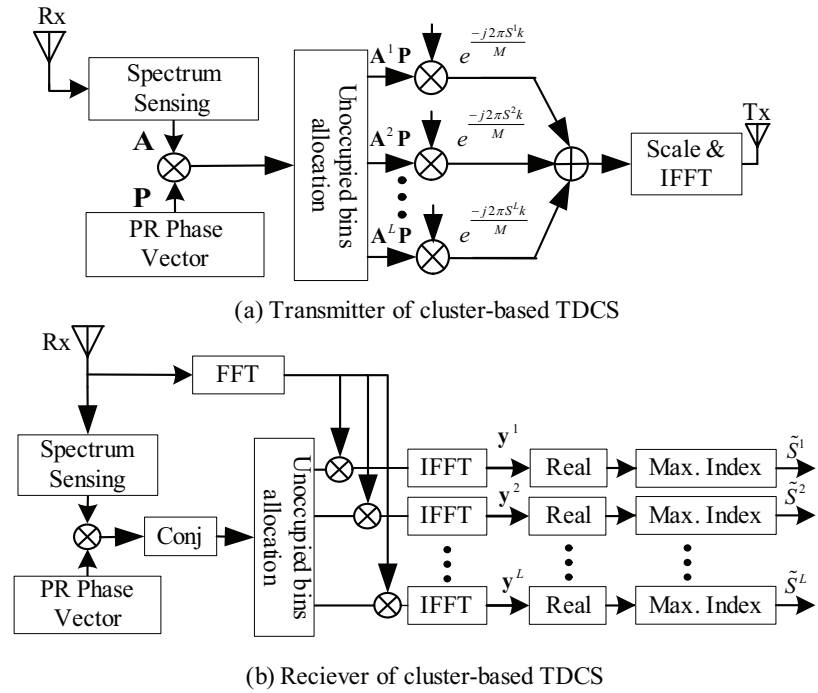}
  \caption{Block diagram of the cluster-based TDCS}
  \label{FIG:TDCSDiagram}
\end{figure}

The FMW representing the $l_{th}$ cluster is generated by performing an IFFT operation on the
scalar product between ${\mathbf{{A}}}^l$ and the PR phase vector $\mathbf{{P}}$, i.e.
$\mathcal {F}^{-1} \left\{ {{\bf{A}}^l  \cdot {\bf{P}}} \right\}$. Then, all FMWs associated with
their corresponding clusters are respectively modulated by the CCSK modulation of an order $M$.
The transmitted waveform $ {\mathbf{{x}}}= \left\{x_0, x_1, \cdots, x_{N-1} \right\}$
generated by the $L$-cluster TDCS is given by
\begin{equation}\label{Eq3}
x_n  = \lambda \sum\limits_{l = 1}^L {\sum\limits_{k = 0}^{N - 1} {A_k^l e^{jm_k } e^{\frac{{ -
j2\pi S^l k}}{M}} e^{\frac{{j2\pi kn}}{N}} } },
\end{equation}
where $\lambda$ is the energy normalization factor given in \eqref{b-FMW} and $S^l \in
\left\{0,1,2,\cdots,M-1 \right\}$ denotes the data symbol carried by the $l_{th}$ cluster. It is
easy to express \eqref{Eq3} into
\begin{equation}\label{Eq4}
x_n  = \lambda \sum\limits_{k = 0}^{N - 1} {\left( {\sum\limits_{l = 1}^L {A_k^l e^{jm_k }
e^{\frac{{ - j2\pi S^l k}}{M}} } } \right)e^{\frac{{j2\pi kn}}{N}} }.
\end{equation}
Therefore, the transmitter of cluster-based TDCS requires only one IFFT operator, as shown in
Fig.\ref{FIG:TDCSDiagram}(a).

After passing through an additive white Gaussian noise (AWGN) channel, the received waveform $
{\mathbf{{r}}}= \left\{r_0, r_1, \cdots, r_{N-1} \right\}$ is
\begin{equation}\label{Eq5}
r_n  = \lambda \sum\limits_{k = 0}^{N - 1} {\left( {\sum\limits_{l = 1}^L {A_k^l e^{jm_k }
e^{\frac{{ - j2\pi S^l k}}{M}} } } \right)e^{\frac{{j2\pi kn}}{N}} }  + w_n,
\end{equation}
where $w_n$ indicates the AWGN noise. Following the CCSK demodulation shown in
Fig.\ref{FIG:TDCSDiagram}(b), the data symbol $S^l$ is recovered by detecting the maximum
correlation output,
\begin{equation}
{\bf{y}}^l = \mathcal {F}^{-1} \left\{ {\mathcal {F} \left\{ {\bf{r}} \right\} \cdot \left(
{{\bf{A}}^l \cdot {\bf{P}}} \right)^* } \right\},
\end{equation}
where $\left( {\bf{A}}^l \cdot {\bf{P}} \right)$ denotes the frequency-domain local reference FMW
associated with the $l_{th}$ cluster. Utilizing the constraint in \eqref{Eq2}, the $\tau _{th}$
element of ${\bf{y}}^l$ is derived as
\begin{equation}\label{Eq7}
y_\tau ^l  = \sum\limits_{p = 0}^{N - 1} { \left( \lambda {\left| {A_p^l } \right|^2 e^{\frac{{ -
j2\pi S^l p}}{M}} } + \left( \sum\limits_{n = 0}^{N - 1} {w_n e^{\frac{{ -j2\pi p n}}{N}} }
\right) {(A_p^l e^{jm_p})^* } \right)e^{\frac{{j2\pi p\tau }}{N}} },
\end{equation}
and the demodulated data symbol $\tilde S^l$ is expressed as
\begin{equation}\label{Eq8}
\tilde S^l  = \arg \mathop {\max }\limits_\tau  \left\{ {\Re \left\{ {y_\tau ^l } \right\}}
\right\} = \arg \mathop {\max }\limits_\tau  \left\{\Re \left\{ {R_\tau ^l  + n_\tau ^l } \right\}
\right\},
\end{equation}
where
\begin{equation}\label{Eq9}
 R_\tau ^l  =   {\lambda \sum\limits_{p = 0}^{N - 1} {\left( {\left| {A_p^l } \right|^2 e^{\frac{{ - j2\pi S^l p}}{M}} } \right)e^{\frac{{j2\pi p\tau }}{N}} } }, \tau=0,1,\cdots,N-1
\end{equation}
denotes the autocorrelation of the $l_{th}$ FMW, and
\begin{equation}
n_\tau ^l  =  \sum\limits_{p = 0}^{N - 1} { \left(  \sum\limits_{n = 0}^{N - 1} {w_n e^{\frac{{
-j2\pi p n}}{N}} }
 {(A_p^l)^* e^{-jm_p}} \right)e^{\frac{{j2\pi p\tau }}{N}} }, \tau=0,1,\cdots,N-1
\end{equation}
denotes the noise obtained by CCSK demodulator. Consequently, the receiver of cluster-based TDCS
is shown in Fig.\ref{FIG:TDCSDiagram}(b), where data symbols corresponding to other clusters can
be recovered by same procedures described above.

Since the cluster-based TDCS can be considered as a group of individual traditional TDCSs where
each one carries $\log _2 \left( M \right)$ bits, the spectrum efficiency of $L$-cluster TDCS is
\begin{equation}\label{EffCl}
\eta _{Cluster} = \frac {L \Delta _f \log _2 (M)}{\gamma W} \text{(bits/s/Hz)},
\end{equation}
where $W$, $\gamma$, and $\Delta_f$ are defined in \eqref{EffTr}.

By comparing \eqref{EffTr} and \eqref{EffCl}, the traditional scheme can be considered as a
special case ($L=1$) of cluster-based TDCS. For a given CCSK modulation order $M$, the spectrum
efficiency is improved only by increasing the spectrum bin spacing $\Delta_f$. However, for the
cluster-based TDCS, two variables in \eqref{EffCl}, $L$ and $\Delta_f$, are associated with the
spectrum efficiency $\eta _{Cluster}$. With the concept of clustering, the cluster-based TDCS is
consisted of a group of individual traditional ones. This arrangement achieves the spectrum
efficiency of $L$-cluster TDCS to be $L$ times of that achieved by the traditional one.

\section{Spectrum bin allocation schemes}
Similar to many communication scenarios, the cluster-based TDCS has a tradeoff between spectrum
and power efficiency. Since the autocorrelation of an ideal FMW has a distinct peak and low
sidelobes, CCSK modulation is a form of $M$-ary signaling over a communication channel
\cite{Dillard}. The lower sidelobes the autocorrelation has, the better BER performance TDCS can
achieve. However, as the number of clusters $L$ increases, the number of unoccupied spectrum bins in each cluster, i.e. $N_C / L$, decreases, leading to high autocorrelation sidelobes. In this
case, BER performance is highly dependent on sidelobes, especially the first few sidelobes.

To better understand the effect of clustering, let us reinvestigate the autocorrelation in
\eqref{Eq9}. Without loss of generality, we may assume the data symbol $S^l=0$ for the $l_{th}$
cluster. According to \eqref{Eq1}, \eqref{Eq9} can be rewritten as
\begin{equation}\label{Eq10}
 R_\tau ^l  =   {\lambda \sum\limits_{p \in \Omega_l^C} {e^{\frac{{j2\pi p\tau }}{N}} } },
 \tau=0,1,\cdots,N-1.
\end{equation}
As a consequence of Cauchy-Schwarz inequality \cite{Cauchy}, for any delay $\tau \neq 0$, $R_0 ^l
\geq R_\tau ^l$ means that $R_0 ^l$ is the autocorrelation mainlobe. Thus, the normalized
sidelobes are expressed as
\begin{equation}
\label{ACnorm} R_{\tau,norm} ^l = \frac{R_\tau ^l}{R_0 ^l} = \frac{L}{N_C} \left( \sum\limits_{p
\in \Omega_l^C} {e^{\frac{{j2\pi p\tau }}{N}} } \right),\tau=1,\cdots,N-1.
\end{equation}

\begin{figure}
\centering
  \includegraphics[width=4.2in]{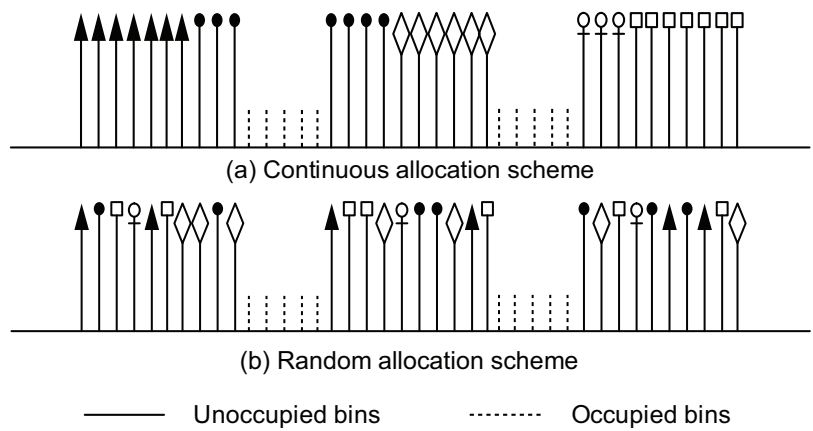}
  \caption{Spectrum bin allocation schemes. (Spectrum bins marked with the same symbol denote the same cluster)}
  \label{FIG:Alloctionscheme}
\end{figure}

From \eqref{ACnorm}, the normalized sidelobes are decided by two factors, the number of clusters
$L$ and the set of unoccupied spectrum bins $\Omega_l^C$. The larger $L$ results in higher
normalized sidelobes, leading to degraded BER performance. For the set $\Omega_l^C$, two spectrum
bin allocation schemes are considered for the cluster-based TDCS, namely continuous and random
allocation schemes shown in Fig.\ref{FIG:Alloctionscheme}.

\begin{figure}
\centering
  \includegraphics[width=4.2in]{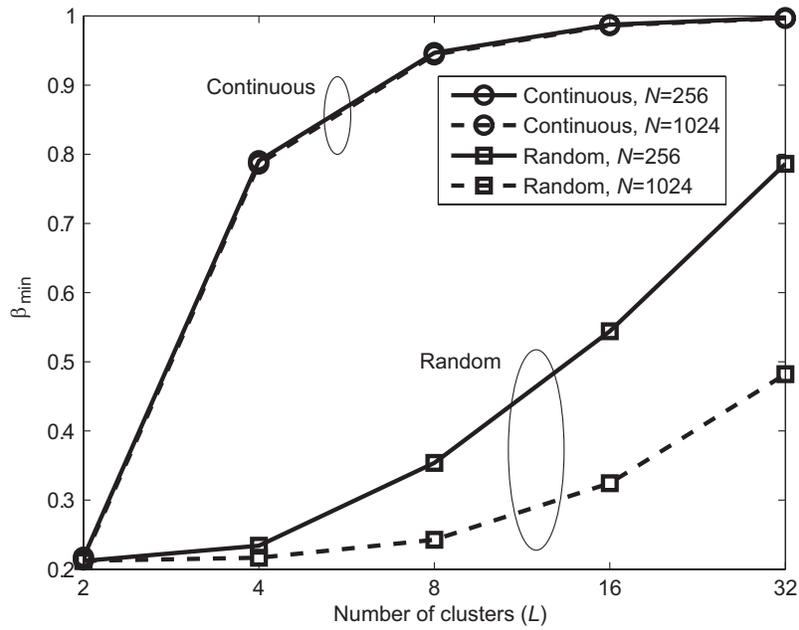}
  \caption{Minimal value of the largest sidelobes obtained from $10^{4}$ Monte-Carlo trials}
  \label{FIG:Sidelobe}
\end{figure}

The objective of all allocation schemes is to minimize sidelobes $\{ R_{\tau,norm}^l, \tau \neq 0 \}$
for all $L$ clusters. In this paper, minimizing the largest sidelobe is considered under the
constraint in \eqref{Eq2}, and the objective function becomes
\begin{eqnarray}
\label{Costf} \beta_{min}  & = & \mathop {\min} \left\{ \mathop {\max}\limits_{l, \tau}
\left\{R_{\tau,norm} ^l, \tau \neq 0 \right\}\right\} \nonumber\\
\text{subject to} & & {\bigcup\limits_{l = 1,2,...,L} {\Omega _l^C }  = \Omega ^C}  \\
&& {\bigcap\limits_{l = 1,2,...,L} {\Omega _l^C }  = \emptyset } \nonumber.
\end{eqnarray}
Utilizing the Stirling approximation \cite{Stirling}, the global search to find out the minimal
$\beta _{\min}$ requires a complexity of
\begin{equation}
\label{Complex}
\prod _{l=0}^{L-1} {\left( \begin{array}{c} \frac{ N_C(L-l)}{L} \\
\frac{N_C}{L}
\end{array} \right)} = \frac{N_C!}{\left( \frac{N_C}{L}! \right)^L} \sim  \left(2 \pi N_C \right)^{\frac{1-L}{2}} \cdot L^{N_C+\frac{L}{2}},
\end{equation}
which means that optimizing the objective function in \eqref{Costf} is a NP-hard problem. As shown
in Fig.\ref{FIG:Sidelobe}, the value of $\beta_{min}$ can also be approximately found by a finite
number of Monte Carlo trials without exhaustive search.

It is obvious that the cluster-based TDCS with the random allocation scheme has a small $\beta_{\min}$
value, and $\beta_{\min}$ gradually increases as the number of clusters $L$ increases. In fact,
with the continuous allocation scheme described in Fig.\ref{FIG:Alloctionscheme}(a), the FMW
corresponding to each cluster has a small total bandwidth, leading to the associated
autocorrelation having high sidelobes. With the random allocation scheme, however, the allocated
unoccupied spectrum bins of each cluster are distributed over almost the entire bandwidth. The
corresponding FMW hence becomes a wide-band signal, leading to low autocorrelation sidelobes. As
the CCSK demodulation in \eqref{Eq8} relies on the FMW with impulse-like autocorrelation
properties, the cluster-based TDCS with random allocation scheme is expected to achieve better BER
performance than that with continuous allocation scheme.

\section{Numerical Results}
To validate the cluster-based TDCS, a scenario of spectrum bandwidth $W = 10$MHz  and $\gamma=3/4$
is considered where the occupied bands are present in the range 2.5$\sim$3.75 MHz and
6.25$\sim$7.5 MHz. In the simulation, we assume $N$ equals $256$ and $1024$, and the CCSK
modulation order equals $N$.

\subsection{Performance in AWGN channel}
\begin{figure}
\centering
  \includegraphics[width=4.2in]{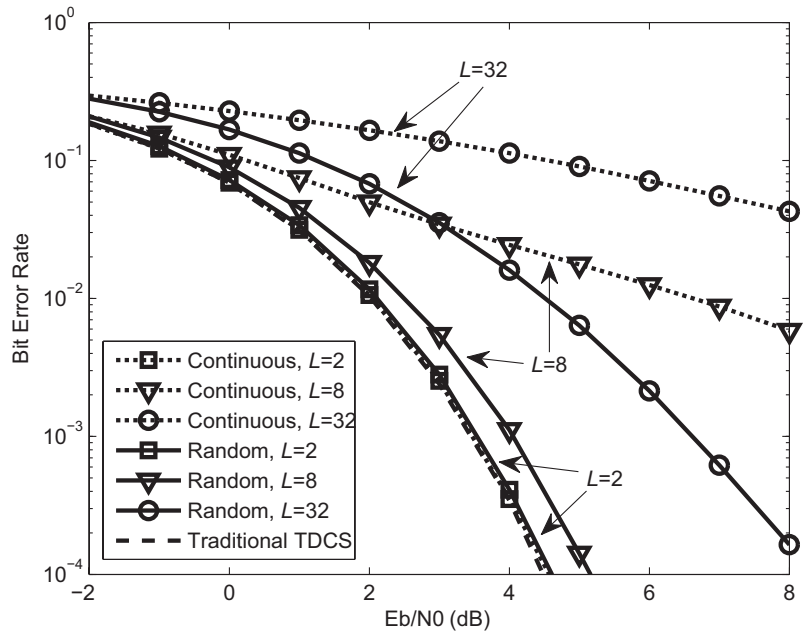}
  \caption{BER performance of the cluster-based TDCS in AWGN channel ($N=256$, and $L=2,8,32$, dashed line: continuous, solid line: random)}
  \label{FIG:BER}
\end{figure}

With the continuous and random allocation schemes, Fig.\ref{FIG:BER} shows the BER performance of
cluster-based TDCS with $N=256$. For a small number of clusters ($L=2$), the TDCS with both
allocation schemes achieves BER performance similar to the traditional one, indicating that the BER performance is not obviously degraded by doubling spectrum efficiency. Although spectrum efficiency is further improved as $L$ increases, however, the proposed TDCS suffers from BER performance degradation. In
particular, the 8-cluster TDCS with random allocation scheme achieves an 8-fold improvement in
spectrum efficiency at the cost of 1dB BER degradation, compared to the traditional scheme.

\begin{figure}
\centering
  \includegraphics[width=4.2in]{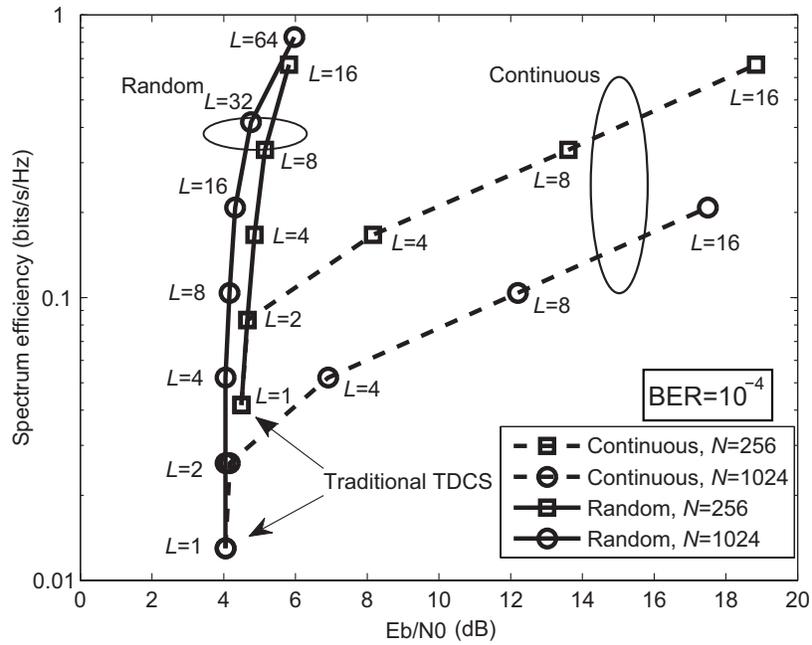}
  \caption{Spectrum and power efficiency of the cluster-based TDCS in AWGN channel ($L=1,2,4,8,16,32,64$ and BER=$10^{-4}$, dashed line: continuous, solid line: random)}  \label{FIG:SpecEff}
\end{figure}

To demonstrate the impact of the number of clusters, $L$, on the system performance,
Fig.\ref{FIG:SpecEff} shows spectrum and power efficiency for $N=256$ and $1024$ in AWGN channel,
i.e., $E_b/N_0$ (dB) required for BER=$10^{-4}$ with $L=1,2,4,8,16,32,64$. In accordance with
analytical results in section IV, the TDCS with random allocation scheme outperforms that with the
continuous scheme in terms of BER performance. For $N=256$, the 8-cluster TDCS with random
allocation scheme achieves a 9dB gain in terms of $E_b/N_0$ compared to that with continuous allocation scheme, because of the lower autocorrelation
sidelobes.

Similar to the results shown in Fig.\ref{FIG:BER}, the cluster-based TDCS suffers from BER degradation when
$L$ is large. Taking an example for $N=1024$, the 8-cluster TDCS with random allocation scheme
requires $E_b/N_0=4.1$dB to achieve BER=$10^{-4}$, whereas the 64-cluster TDCS requires
$E_b/N_0=6.1$dB. This observation indicates that spectrum efficiency is increased from $0.104$ to
$0.833$ (bits/s/Hz) with a penalty of $2$dB in terms of $E_b/N_0$ to achieve BER=$10^{-4}$.
Therefore, for practical scenarios, TDCS should choose a suitable value of $L$ to achieve a
desirable tradeoff between the spectrum and power efficiency requirement.

\subsection{Performance in multipath fading channel}
Let us discuss the performance of coded TDCS in multipath fading channel (COST207RAx6 channel in
\cite{Cost207}). A convolution channel code with a coding rate $1/2$ is considered. To combat the
effects from the multipath fading channel, the length $1/4$ cyclic prefix and a minimal mean
square error (MMSE) equalizer are simulated.

\begin{figure}
\centering
  \includegraphics[width=4.2in]{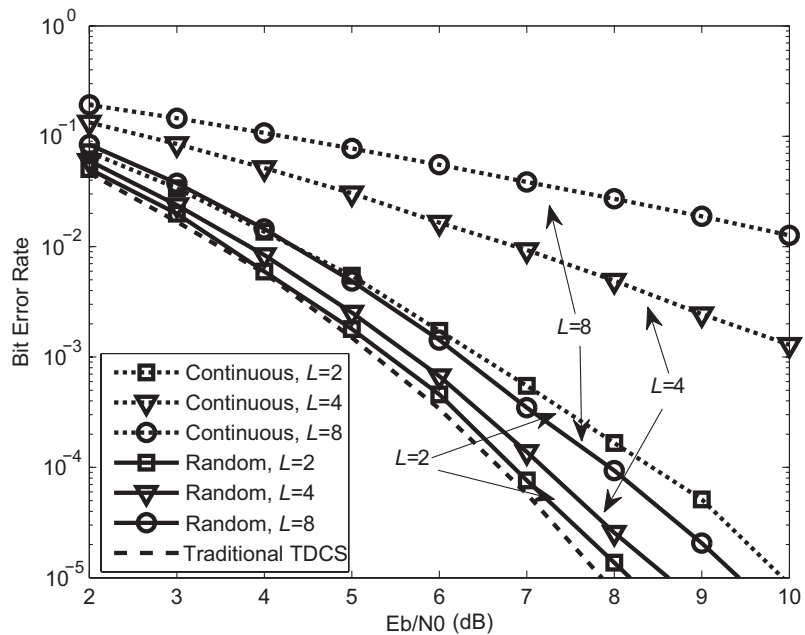}
  \caption{BER performance of the coded cluster-based TDCS in multipath fading channel ($N=256$, $L=2,4,8$, dashed line: continuous, solid line: random) }
  \label{FIG:BERM}
\end{figure}

Fig.\ref{FIG:BERM} shows that the TDCS with both allocation schemes achieves degraded BER
performance when $L$ increases. It is also interesting to observe that, for $L=2$, the TDCS with
random allocation scheme is superior to that with the continuous scheme, in contrast to the AWGN
channel case where both allocation schemes achieve similar BER performance. Compared to the
continuous allocation scheme, each FMW associated with random allocation scheme spreads over a
wider spectrum. Therefore, the cluster-based TDCS with random allocation scheme achieves better
BER performance than that with continuous scheme, due to a larger degree of frequency diversity in
multipath fading channel \cite{Papproth}.

\begin{figure}
\centering
  \includegraphics[width=4.2in]{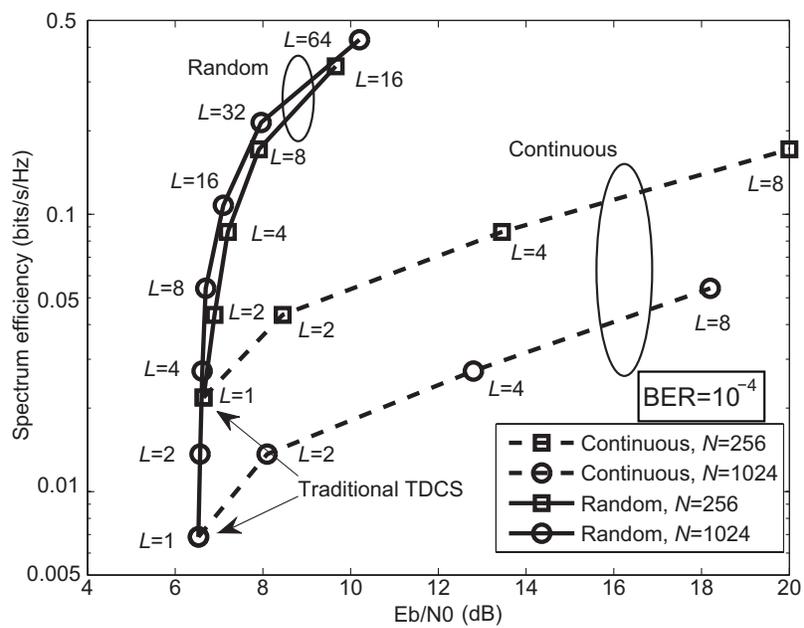}
  \caption{Spectrum and power efficiency of the coded cluster-based TDCS in multipath fading channel ($L=1,2,4,8,16,32,64$, BER=$10^{-4}$, dashed line: continuous, solid line: random)}
  \label{FIG:SpecEffM}
\end{figure}

Fig.\ref{FIG:SpecEffM} illustrates the spectrum and power efficiency in multipath fading channel,
where the TDCS with random allocation scheme still achieves better spectrum and power efficiency,
compared to that with continuous allocation scheme. Furthermore, the cluster-based TDCS suffers
from obvious BER degradation when $L$ exceeds a certain threshold value. According to the
simulation results, the number of clusters should be $L \leq 4$ for $N=256$ and $L \leq 16$ for
$N=1024$.

According to the system performances in AWGN and multipath fading channels, we make the following
remarks.
\begin{itemize}
  \item To achieve better spectrum and power efficiency, the cluster-based TDCS should adopt
  the random allocation scheme. Since the randomly allocated bins
  are distributed over almost the entire bandwidth, the generated FMW has low autocorrelation sidelobes leading to the robust
  BER performance. Furthermore, in multipath fading channel, the proposed TDCS with random allocation scheme achieves more robust BER performance due to a
larger degree of frequency diversity, compared to that with continuous allocation scheme.
  \item Due to the tradeoff between spectrum and power efficiency, the cluster-based TDCS cannot unlimitedly
  increase spectrum efficiency. When $L$ exceeds the specific threshold value, such as $L = 4$ for $N=256$ and $L = 16$ for $N=1024$, BER performance rapidly degrades. This
  observation provides a quick rule of thumb for designing the cluster-based TDCS.
  \item The traditional scheme can be considered as a special case ($L=1$) of
cluster-based TDCS, where only a large value of spectrum bin spacing $\Delta_f$ can improve
spectrum efficiency. Under the constraint of a small value $\Delta_f$ for robust BER performance,
this inherent tradeoff makes it difficult to achieve robust BER performance and high spectrum
efficiency simultaneously. Fortunately, with the concept of randomly clustering, the cluster-based
TDCS with a smaller $\Delta_f$ achieves higher spectrum efficiency and more robust BER
performance.
\end{itemize}

\section{Conclusion}
In this paper, we have proposed a cluster-based TDCS framework to improve spectrum efficiency by
dividing all unoccupied spectrum bins into clusters. Among various schemes of spectrum bin spacing
and allocation, analytical and simulation results show the proposed TDCS with random allocation
scheme achieves higher spectrum efficiency and more robustness against BER performance
degradation, compared to that with continuous allocation scheme. Furthermore, different from
previously reported conclusions in the literature, the cluster-based TDCS should configure a
smaller spectrum bin spacing $\Delta _f$ to achieve higher spectrum efficiency and more robust BER
performance.

\newpage

\end{document}